\documentclass[runningheads]{llncs}
\usepackage[T1]{fontenc}
\usepackage{graphicx}
\usepackage{cite}
\usepackage{amsmath,amssymb,amsfonts}
\usepackage{algorithm}
\usepackage{algpseudocode}
\usepackage{textcomp}
\usepackage{booktabs}
\usepackage{url}
\usepackage{array}
\usepackage{hyperref}

\begin{document}
    
\title{Deep Clustering for Blood Cell Classification and Quantification}
\author{Mihaela Macarie-Ancau\inst{1}\orcidID{0009-0002-5743-4945} \and
Adrian Groza\inst{1}\orcidID{0000-0003-0143-5631}}

\authorrunning{M. Macarie-Ancau and A. Groza}

\institute{Department of Computer Science, Technical University of Cluj-Napoca, Romania \\
European University of Technology (EUt+), European Union\\
\email{Macarie.Io.Mihaela@cs.utcluj.ro, Adrian.Groza@cs.utcluj.ro} \\
}

\maketitle

\begin{abstract}
Accurate classification of blood cells plays a key role in improving automated blood analysis for both medical and veterinary applications. This work presents a two-stage deep clustering method for classifying blood cells from high-dimensional signal data. In the first stage, red blood cells (RBCs) and platelets (PLTs) are separated using a combination of an improved autoencoder and the IDEC algorithm. The second stage further classifies RBC subtypes, pure RBCs, reticulocytes, and clumped RBCs, through a variational deep embedding (VaDE) approach. Due to the lack of detailed cell-level labels, soft classification probabilities are generated from sample-level data to approximate the true distributions. The aim is to contribute to the development of low-cost, automated blood analysis systems suitable for veterinary and biomedical use. Initial results indicate this method shows promise in effectively distinguishing different blood cell populations, even with limited supervision.
\end{abstract}


\section{Introduction}
From the smallest whisper of blood, microscopic analysis unveils a vast wealth of cellular data, fundamentally reshaping our understanding of an organism's intricate dance between health and disease. 
This analysis plays an important role in the detection of infections, the diagnosis of hematological disorders, and in monitoring immune function, and guiding treatment.

Yet, the complexity of this data also poses ongoing challenges. 
Only through expert interpretation and high-speed equipment that is accurate and sensitive to abnormalities can its full value be achieved. Current technologies are often insufficient when extrapolated from human diagnostics to other species. Blood morphological difference among animals makes the development of traditional rule-based models difficult without extensive pre-analysis, feature engineering, or species-specific tuning. 
Such a challenge demands well-curated data and medical expert validation, especially for uncommon or inconspicuous cell populations. Also, throughput capacity, variability, and multi-parameter integration are even more critical in multi-veterinary practice. These barriers must be overcome along the path of creating scalable, accurate, and medically significant diagnostic technology across species.
By addressing these challenges, one can enhance diagnostic understanding and clinical efficacy.
We propose here a two-stage deep clustering method for classifying blood cells from high-dimensional signal data.
\section{Technical instrumentation}

\subsection{How a blood count instrument works}

\begin{figure}
    \centering
    \includegraphics[width=1\textwidth]{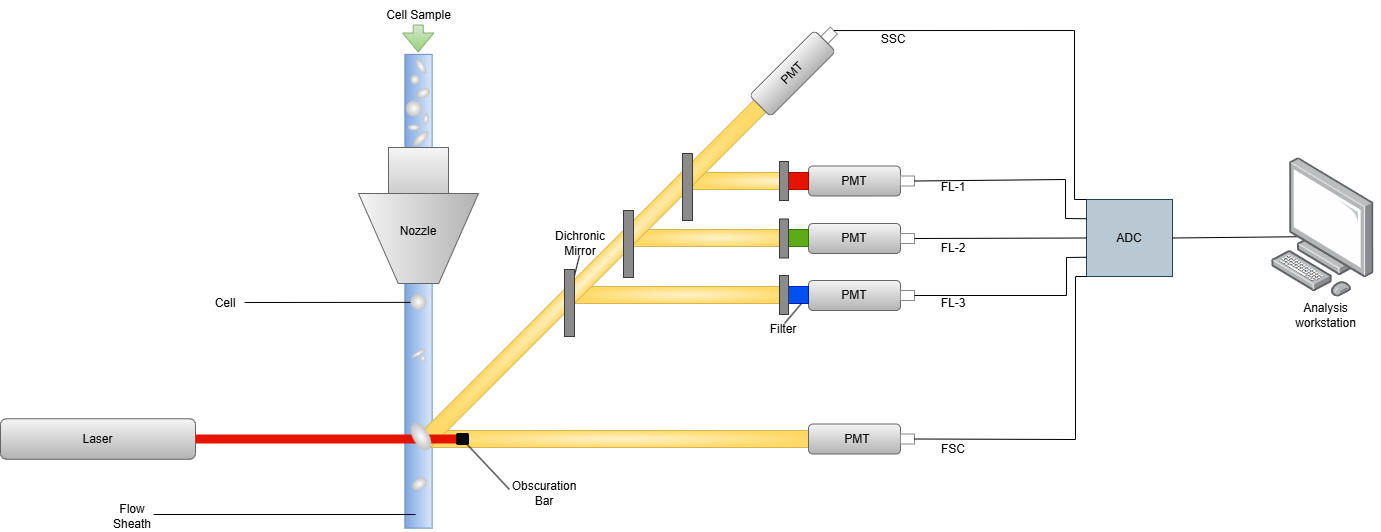}
    \caption{Internal representation of how cells are identified}
    \label{fig:cells}
\end{figure}

Hematology analyzers, i.e. blood count instruments, receive a sample of blood, mixed with anticoagulant to separate possible clumps, and classify various cells by analyzing their electrical, optical, or laser-based signal profiles.
That is, Complete Blood Counts (CBC).
We relied on  cytometry-based instrument, where a laser beam is directed at cells as they flow in a single line through a small tube.
Multiple channels are used, each capturing different signal features (Figure~\ref{fig:cells}).

As each cell interacts with the laser, the detectors capture light scattered at various angles and through different color filters: (i) Forward Scatter (FSC), to determine the cell size and 
(ii) Side Scatter (SSC) to determine the opacity or internal complexity of a cell.
The resulting data are then plotted in the form of time series of pulses, which provide visual representations of the composition and characteristics of the cells. Data can include distinctions among different types of blood cells, such as red blood cells, white blood cells, and platelets. 
The precision of this data collection is crucial to obtain reliable counts and classifications of blood components. 

\subsection{Limitations of traditional blood count instruments}
Current instruments rely on rule-based classification and traditional algorithms and they often struggle when ambiguous or atypical samples, such as: 
(i) \textit{reticulocytes} (immature red blood cells),
(ii) \textit{coincidences} (red blood cell clumps or aggregates), and
(iii) \textit{platelets} with abnormal morphology.

In addition, these instruments are calibrated for human blood. Given the complexity of blood samples, particularly in the veterinary context, where patients come from different species, one anticoagulant can work differently for two blood samples. Even if the initial goal is met, to get rid of cell clumps, external factors (e.g. time) can interfere and can lead the algorithm inside the instrument to produce ambiguous results when samples are given from other species. 

\begin{figure}
    \centering
    \includegraphics[width=1\textwidth]{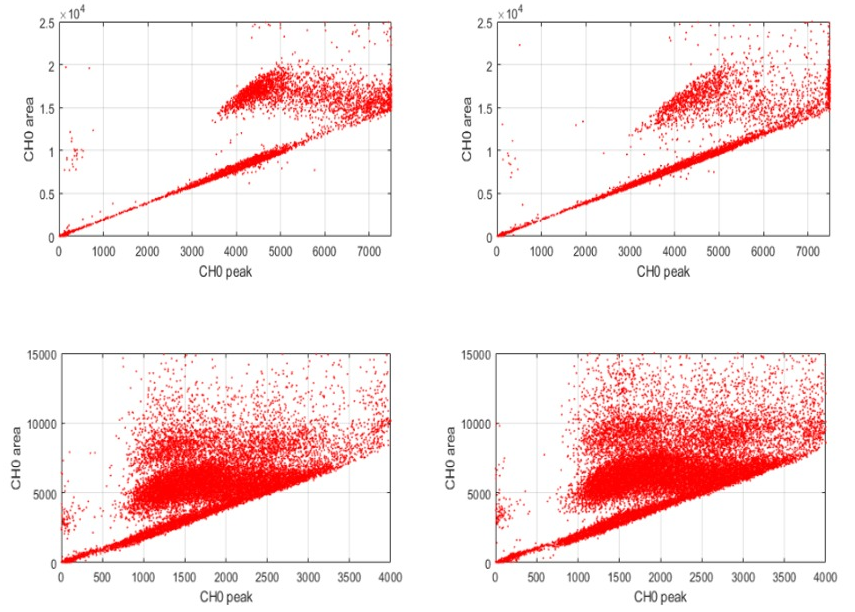}
    \caption{Clusters of cells from human (top) vs. cat blood (bottom)}
    \label{fig:cat}
\end{figure}

In Figure~\ref{fig:cat}, the clusters from the human blood sample are much easier to be spotted than the ones from a cat blood sample. The platelets, which are the smallest ones, should be in a proportion of 5\% and
the rest are red blood cells (the white cells are washed out before counting). 
The red blood cells fill the rest of the plot, and can be in three categories: 
(1) \textit{regular} mature red blood cells
(2) \textit{reticulocytes}, young red blood cells created less than 24 hours ago, and
(3) \textit{coincidences}, i.e. two or more red blood cells clustered together.

Differently, the cat blood cells have a much higher chance to create doublets or triplets, making it harder to count them. 
We can use general percentages as weights for each category, but that can end up bad when there is a problem and we do not see it. We need to have a precise count and classification of each cell.

\subsection{What do we have available?}
The instrument processes the blood sample by introducing a reagent that de-coagulates the blood cells and filters out white blood cells (WBCs), allowing only red cells (RBCs) and platelets (PLTs) to flow through the measurement tube. 
Ideally, this leads to a clean separation of the cell types without any aggregates.

In practice, timing plays a critical role. 
Between the moment the reagent is added and when the sample passes through the tube, a variable delay may occur. During this interval, cells can begin to re-coagulate, forming aggregates or clumps known as coincidences.
These coincidences complicate the analysis since they appear as merged signals that do not correspond to individual cells. 

Off-the-shelf deep clustering algorithms could help in several ways: 
(i) giving us values to use as a ground truth for each cell;
(ii) fine-tuning a model to use for instrument computation instead of the current algorithm;
(iii) training a model to find the percentage of reticulocytes which is helpful for diagnose anemia;
(iv) training a model to help us identify and classify blood cells for different animals instead of using different algorithms which require a long time for optimization.

\subsection{Model selection for predicting blood count}

Based on the dataset consisting of 11 dimensions, two primary approaches for selecting a suitable model are: (i) using pre-built models or (ii) constructing a custom artificial neural network. 
The decision depends on the desired performance or interpretability of the model.

\textit{Pre-Built Models for Tabular Data}
are broadly effective, but not not optimally suited to our task. 
The complexity of bio signals and our goal of uncovering subtle, unlabeled patterns in an evolving, unsupervised environment demand a specialized approach. Accordingly, this study investigates deep clustering methods and custom neural architectures that address these requirements.

\textit{Custom Neural Networks} provide a promising direction towards more flexibility and the ability to dynamically learn required features from sophisticated data such as those in our study. It is possible to customize these networks to the particular complexities of the data, and this should provide deeper insight and improved performance over the conventional techniques.

We experimented here with deep clustering and generative methods, which are highly suitable for unsupervised and weakly supervised tasks, especially in scenarios with limited labeled data, such as blood cell classification. The specific approaches investigated include:
\begin{itemize}
    \item \textit{N2D (Not Too Deep) Clustering\cite{McConville2021}:} This approach combines deep representation learning (typically via an autoencoder) with subsequent manifold learning on the obtained embedding, followed by a traditional clustering algorithm applied to the learned manifold.
    \item \textit{Vanilla Autoencoders (AE)\cite{Bank2021}:} Neural networks used to learn efficient dimensionality reduction by reconstructing their input. The learned latent space can then be used for clustering.
    \item \textit{Variational Autoencoders (VAE)~\cite{Kingma2022}:} A generative extension of AEs that learns a probabilistic latent space, useful for generating new data and for clustering.
    \item \textit{DEC (Deep Embedded Clustering)\cite{Xie2016}:} A method that simultaneously learns feature representations and cluster assignments using a deep neural network.
    \item \textit{IDEC (Improved DEC)\cite{Guo2017}:} An extension of DEC that incorporates an autoencoder reconstruction loss to preserve local structure in the data, improving clustering performance.
    \item \textit{VaDE (Variational Deep Embedding)\cite{Jiang2017}:} A generative clustering approach that models the data distribution with a Gaussian Mixture Model (GMM) fitted over the latent space learned by a VAE.
\end{itemize}
Such models combine feature learning and clustering in an end-to-end manner, and they are suitable for unsupervised or weakly supervised tasks like our blood cell classification, where there are no per cell labels.

\section{Method}

A preliminary investigation with a basic Improved Deep Embedded Clustering (IDEC) approach guided the creation of a more advanced, better fitted on our case, two-stage deep learning methodology.

Current algorithm parameters, such as the number of clusters or latent dimensionality, were selected based on experimenting and observing their performance and domain knowledge. While this approach had promising results, it introduces subjectivity and may not work the same on different datasets. One improvement comes from automated hyperparameter optimization. 
Techniques such as Bayesian optimization, grid/random search, and evolutionary algorithms can be used to systematically explore the configuration space and identify parameter combinations that yield better clustering performance, reconstruction accuracy, and predictive stability. This could result in a more robust and generalizable model, especially when adapting the pipeline to new data sources, blood analyzers, or animal species.

\subsection{Stage 1 - Latent representation and initial clustering}
Stage 1 involves creating an autoencoder and training it to obtain a more discriminative latent space representation of the input cell data, in the scope of enhancing data separability. Various autoencoder architectures have been experimented on, including different numbers of layers, neuron counts per layer, and activation functions.

The latent representations from the encoder are passed to an Improved Deep Embedded Clustering (IDEC) module, which simultaneously performs clustering and reconstruction. A regression head is also added to predict the relative proportions of red blood cells (RBCs) and platelets (PLTs) in each sample, encouraging alignment between clusters and cell types. This stage forms the backbone of the pipeline by cleanly separating RBCs from PLTs in an unsupervised yet biologically meaningful manner.

\subsection{Stage 2 – RBC subclassification via VaDE}
Stage 2 focuses exclusively on the RBCs identified in Stage 1. They are passed to a Variational Deep Embedding (VaDE) algorithm, enabling a more specialized and probabilistic RBC classification into three biologically distinct subtypes: clumps, reticulocytes, and normal RBCs. This stage uses the power of variational inference for uncertainty modeling and soft clustering, refining the overall system granularity and accuracy.

\begin{figure}
    \centering
    \includegraphics[width=1\textwidth]{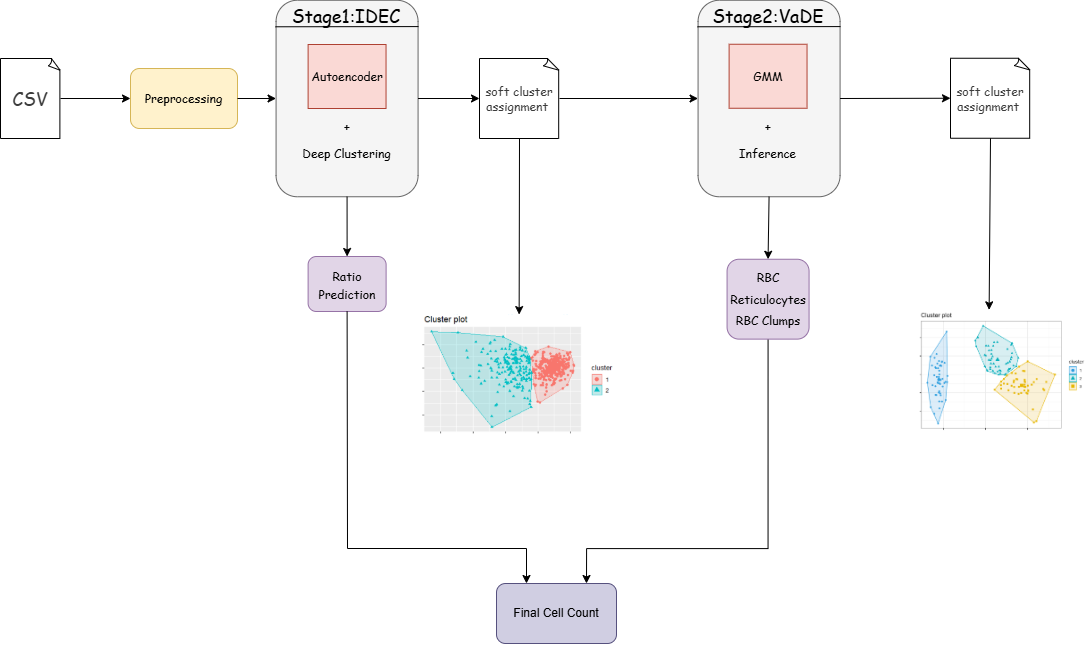}
    \caption{System architecture}
    \label{fig:sys_arch}
\end{figure}

\subsection{Data acquisition and preprocessing}

Raw data is obtained from blood count instruments that record multiple characteristics of cellular signals across different channels. Rather than capturing raw waveform data, each detected event (e.g., an individual cell or cell cluster) is represented as a tabular feature vector. These vectors include derived features such as peak values, area under the curve, and timing features for each channel.

Each sample in the dataset corresponds to a single blood measurement and contains tens of thousands of such feature vectors. The dataset is structured so that each CSV file contains signals from one biological sample, with features extracted from multiple measurement channels.

The data for this research was created with the Aquarius instrument from Diatron \cite{DiatronAquarius3}. It is structured in multiple CSV files, where each file corresponds to one individual blood sample. Each row in a file corresponds to one individual laser-detectable event, and is characterized by 11 different features. These are based on the five channels of the device: four of these channels, at varying angles and with varying optical filters, measure peak signal and area. The fifth channel, being oriented crosswise to the laser beam, having heightened sensitivity to cell size, besides its peak and area values, yields an additional, temporal measurement, for a total of 11 distinct features per event. Each event counted can reflect an individual cell type, such as a red blood cell (RBC), a platelet (PLT), or a reticulocyte, or it may reflect a coincidence event, in which a cluster of cells (notably RBCs) is counted as a single event. The clumping may result if cells re-aggregate over time following addition of an anticoagulant or if they did not get separated on the first hand. The sample preparation involved mixing a specified volume of blood with a proprietary anticoagulant designed to optimize cell separation, with each sample analyzed resulting in a mixture containing about 150000 cells, represented as signals in the CSV file produced, delivering full data for analysis.

Preprocessing involved feature scaling and normalization, specifically using standardization, to ensure numerical stability and consistency across samples, which is an essential prerequisite for our deep learning approach. Observing a high degree of intra-sample homogeneity, we adopted a strategy centered on processing representative batches drawn from each CSV file. This approach operated under the assumption that sufficiently large, random subsets would capture the cell population distributions characteristic of the full sample, much like estimating the density of a forest from a well-chosen patch. Beyond this conceptual rationale, batch-wise processing also addressed practical challenges such as memory constraints and computational efficiency, common limitations in deep clustering applications. Depending on the requirements of specific training phases or models, data was either handled in these representative batches or processed as complete per-file datasets to retain broader context when needed.

\subsection{Output and visualization}

The system outputs the number of cells for a specific sample, identifying populations such as red blood cells and platelets, enabling clinical interpretation and quality control.
Visualization tools generate scatter plots of latent embeddings colored by cluster assignment, facilitating visual inspection of cluster quality and detection of coincident events or artifacts.

\section{Running Experiments}

\subsection{Autoencoder architecture experiments}
Stage 1 begins with learning a latent representation of blood cell data using an autoencoder. Several architectures were evaluated by varying the number of encoding layers, the number of neurons per layer and the activation functions.

Initially, a minimal design with only two linear layers was tested, directly mapping the input features to the latent dimension. While simple, this model showed limited capacity to extract hierarchical and abstract features from the complex multi-channel blood cell data. To address this, an intermediate hidden layer was added, resulting in a three-layer encoder:

Different sizes of the hidden layer were experimented with, typically ranging from 64 to 512 neurons. Larger layer sizes offered increased model capacity but risked overfitting, while smaller sizes improved generalization but reduced representational power. The final chosen size balanced these trade-offs based on clustering performance and reconstruction error.

\begin{figure}
    \centering
    \includegraphics[width=0.8\textwidth]{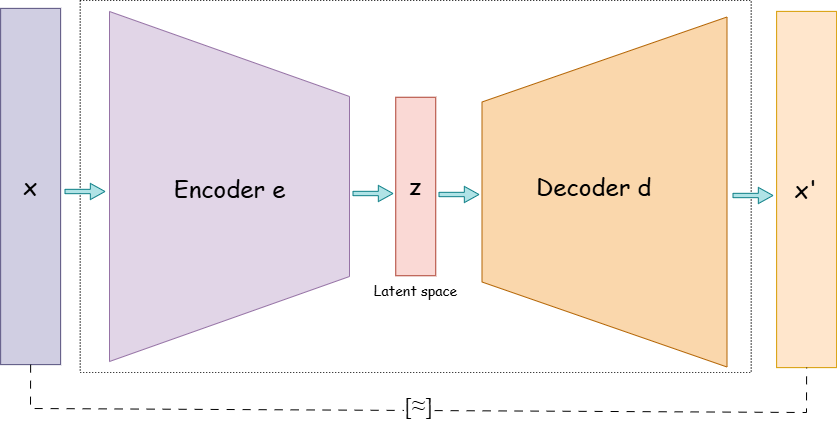}
    \caption{Autoencoder architecture design}
    \label{fig:ae_arch}
\end{figure}

ReLU (Rectified Linear Unit) was chosen as the activation function between layers because of its proven effectiveness in deep learning tasks~\cite{Agarap2019}. ReLU introduces non-linearity such that the network can learn complex patterns. It is computationally cheap and avoids the vanishing gradient problem, thereby cutting down training time and stabilizing it. In addition, sparse activation (gives output zero for negative inputs) encourages the network to learn compact and disentangled features, beneficial for clustering tasks like the segregation of PLTs and RBCs.

\subsection{IDEC training, batch strategy, hyperparameter tuning}
\begin{figure}
    \centering
    \includegraphics[width=1\textwidth]{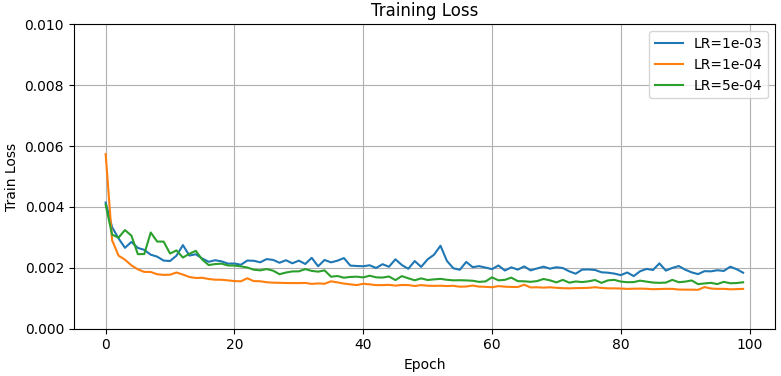}
    \caption{IDEC loss comparison}
    \label{fig:idec_loss}
\end{figure}

Following autoencoder training, the latent representations were fed into the IDEC algorithm, which performs joint clustering and reconstruction, with an added regression head to predict RBC and PLT ratios per blood sample. To preserve sample integrity and reduce variability, mini-batching was applied within individual CSV files, treating each as a separate blood given the homogeneity of the data. 
To assess convergence stability and clustering performance, various learning rates were tested during IDEC training. Specifically, learning rates of \(1 \times 10^{-3}\), \(1 \times 10^{-4}\), and \(5 \times 10^{-4}\) were evaluated as can be seen in Figure~\ref{fig:idec_loss}. These values provided reasonably stable training dynamics and interpretable cluster structures. Higher learning rates such as \(1 \times 10^{-2}\) and \(5 \times 10^{-2}\) were also tested but led to highly unstable training, with frequent spikes in the loss components and poor convergence. As a result, these configurations produced unreliable clustering results and were excluded from the plots.

The clustering soft assignments are computed via the Student’s t-distribution:
\[
q_{ij} = \frac{
    \left(1 + \frac{\| z_i - \mu_j \|^2}{\alpha} \right)^{-\frac{\alpha+1}{2}}
}{
    \sum_{j'} \left(1 + \frac{\| z_i - \mu_{j'} \|^2}{\alpha} \right)^{-\frac{\alpha+1}{2}}
}
\]

\begin{figure}
    \centering
    \includegraphics[width=1\textwidth]{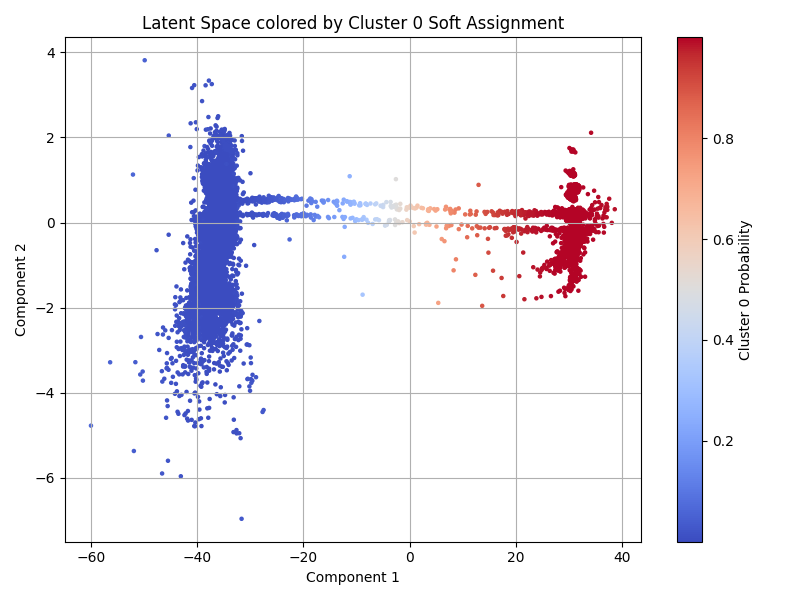}
    \caption{Clustering with IDEC output}
    \label{fig:idec_clusters}
\end{figure}

Stable training and meaningful cluster separation were achieved when balancing losses appropriately as can be seen in the Figure ~\ref{fig:idec_clusters}. Regression head improved alignment between clusters and actual cell type proportions.

\begin{figure}
    \centering
    \includegraphics[width=1\textwidth]{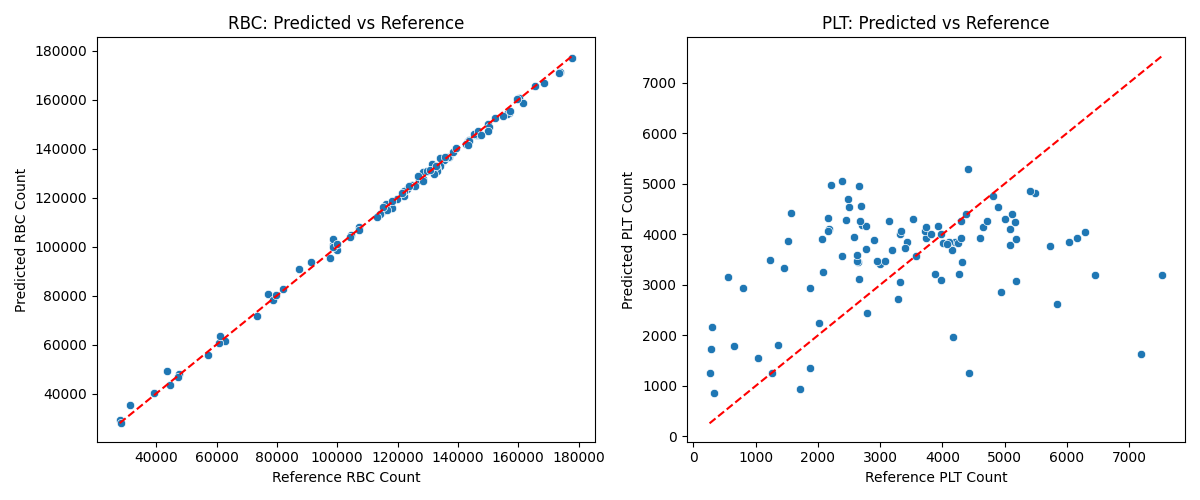}
    \caption{Visualization of IDEC predictions}
    \label{fig:idec_scatter}
\end{figure}

Considering the huge number of cells in one sample and the average ratio between them to be 95\% RBC to 5\% PLT, the RBC prediction got a very good \(R^{2}\) score of 0.998, meaning that the algorithm explains nearly all the variation in the red blood cell counts, indicating excellent accuracy and reliability as can be seen in Figure~\ref{fig:idec_scatter}. 
This level of predictive accuracy not only testifies to the quality of the deep learning approach but also to the application potential for medical and veterinary clinical environments. 

On the other hand the prediction of PLT is much more fragile. For example, a misclassification of just a couple hundred cells, which might seem minuscule looking at the total RBC count, can impact significantly the PLT count. In some cases this error can even double the number of platelets. This happens because the platelets are much less abundant overall, and the model has not enough information to properly learn their configuration. Additionally, since the model predicts the ratios jointly to sum to 1, the PLT and RBC ratios are interdependent, meaning that an error in one of them is clearly an error in the other as well. However, an average error in PLT ratio is insignificant to the RBC ratio, while one in RBC ratio has huge impact for the resulting PLT count. 

To address this, future work will focus on strategies for imbalanced learning:
\begin{itemize}
    \item \textit{Focal Loss:} A dynamic loss function that down-weights the majority class examples and focuses learning on the minority class samples, which would encourage the model to pay greater attention to PLT features.
    \item \textit{Class-Balanced Loss:} This loss formulation re-weights training examples based on the inverse frequency of each class, helping to counteract the imbalance in sample distribution.
    \item \textit{Synthetic Oversampling:} Techniques such as SMOTE (Synthetic Minority Over-sampling Technique)  \cite{SMOTE2011}  can be used to generate synthetic examples of minority classes by interpolating between existing samples. This approach can help balance class distributions and improve the ability to learn under-represented patterns without significantly increasing the risk of overfitting.
\end{itemize}

Experimenting with these approaches could lead to more robust PLT predictions, even in high-imbalance scenarios commonly encountered in hematological datasets.
Nonetheless, the very good RBC prediction provides a solid foundation on which to proceed. Additionally, the classification of 109 CSV files (each containing about 150k cells) was completed in approximately 60.6 seconds, demonstrating that the deep learning pipeline is also efficient enough to handle large datasets in a practical timeframe. 

\subsection{RBC subclassification with VaDE}

Stage 2 uses the Variational Deep Embedding (VaDE) model to further subclassify RBCs identified in Stage 1 into clumps, reticulocytes, and normal RBCs. VaDE jointly trains a Variational Autoencoder and a Gaussian Mixture Model, modeling the latent space as a mixture of Gaussian components instead of a single Gaussian. This enables probabilistic clustering with uncertainty estimation.

\begin{figure}
    \centering
    \includegraphics[width=0.8\textwidth]{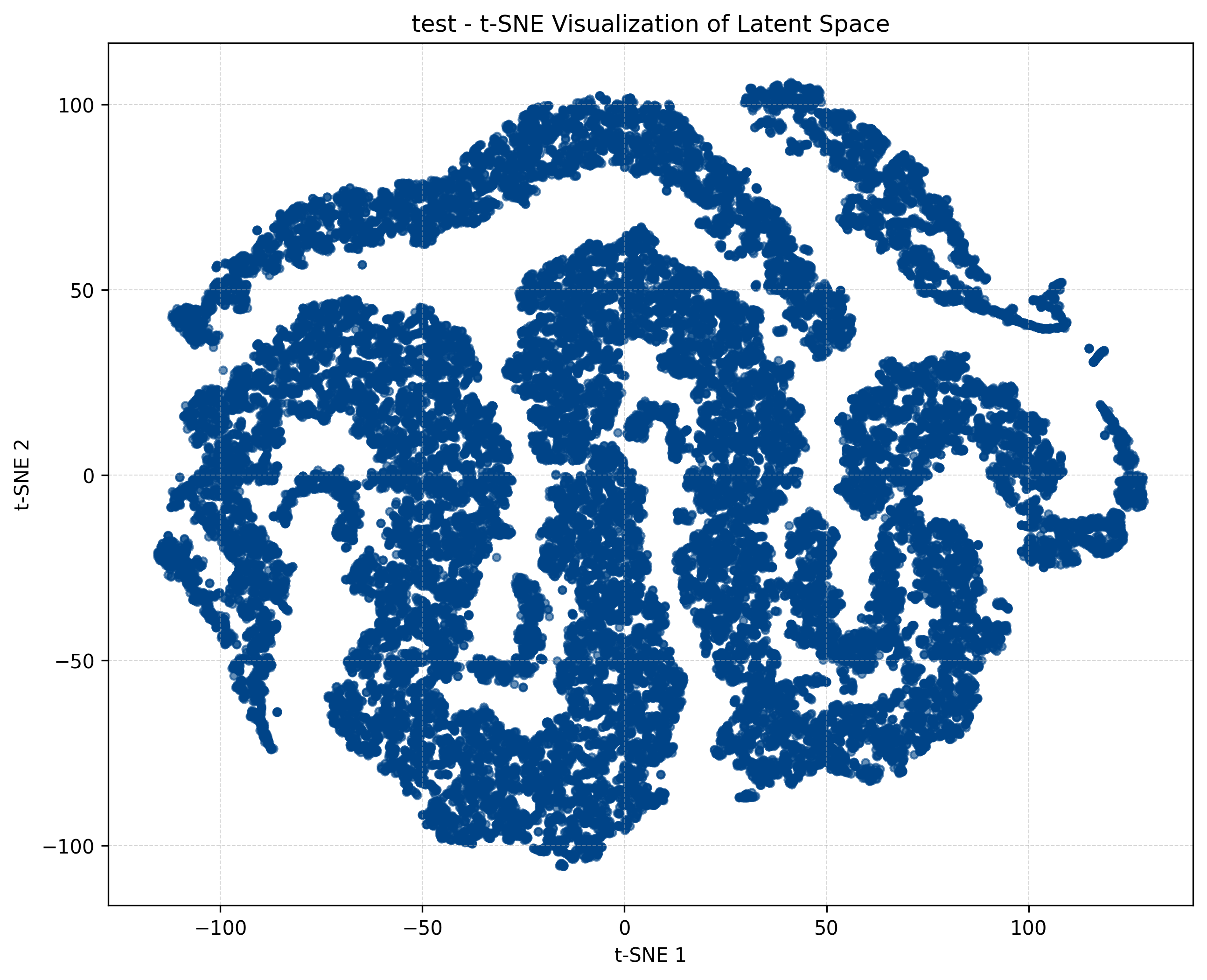}
    \caption{Visualization of VaDE applied over the Latent Space}
    \label{fig:vade_plot}
\end{figure}

As a clustering methodology note, 
the model performs unsupervised clustering natively within the latent space without any outside validation. As can be seen in the Figure \ref{fig:vade_plot} the plot illustrates: 
(1) \textit{Raw Latent Space Projection}: Points are encoded according to their actual latent coordinates from the VaDE encoder;
(2) \textit{No External Validation}: No ground-truth labels or external clustering metrics were used. This decision was made due to the absence of reference annotations and in order to preserve and observe the  natural output without interference.

This setup was designed as an exploratory experiment to examine VaDE’s capacity to learn meaningful structure from the data in an unsupervised manner. The observed cluster separations arise solely from the learned Gaussian mixture model within the latent space and reflect the internal representation of the data distribution, without relying on predefined labels
However, without expert-annotated subtype labels, it remains challenging to objectively assess the accuracy and biological relevance of the discovered clusters. To strengthen the reliability of subtype assignments, future work will incorporate validation mechanisms, such as clinical reference comparisons or human-in-the-loop assessments, to align the resulted unsupervised outputs with domain expertise.

\textbf{Code availability.} The source code is available \href{https://github.com/Mikaelis15/DeepFlow}{Here}.
The demo application can be accessed \href{https://drive.google.com/drive/folders/18Rif6adtXzgycEJJl2LF8H3Vr5P63Klx?usp=drive_link}{Here}. 
These resources include the a video of the application running and some important pieces of code.

\section{Discussion and related work}
The application of ML to hematological analysis is a growing field of study, driven by the potential for enhanced diagnostics precision, throughput, and less reliance on labor-intensive manual workflows. Supervised deep learning has proven to be effective, particularly in the context of image-based analysis. For instance, researchers have shown that it is possible to accurately classify human white blood cells using stain-free imaging flow cytometry data, highlighting how these models can learn meaningful features from complex cell morphology without relying on traditional staining methods~\cite{Lippeveld2020}. This achievement highlights a major trend: the move towards automated systems that can extract clinically relevant information directly from raw instrumental data.
However, one significant bottleneck for supervised learning is its requirement for big, painstakingly annotated datasets. In many biomedical applications, like examining data from high-throughput blood counters, it is often not feasible or very expensive to obtain precise, cell-by-cell labels. This has instigated active interest in unsupervised learning methods, which have the potential to detect intristic patterns in data without explicit labels. One of the most promising paradigms among them is deep clustering, since it combines the strength of deep neural networks in learning rich, hierarchical features with the task of data clustering. 

One of the early significant contributions in this area is Deep Embedded Clustering (DEC), proposed by Xie et al.~\cite{Xie2016}. 
The main novelty of DEC was to use a ML method that learns a mapping from the data space to a low-dimensional feature space and, at the same time, updates cluster assignments in that space. This is achieved by adding a clustering specific loss function that encourages the learned embeddings to form clusters around cluster centroids. Innovative as it is, one potential pitfall of the original DEC formulation is that the clustering loss can, during training, increasingly distort the learned feature space, undermining the autoencoder's ability to preserve the local structure of the data. 
To alleviate this, Guo et al. introduced Improved Deep Embedding Clustering (IDEC), a simple yet effective variant of DEC approach \cite{Guo2017}. IDEC elegantly resolves the feature space distortion problem by incorporating the autoencoder's reconstruction loss into the total objective function, where the feature learning components are trained jointly. In this manner, the representation in the latent space is ensured to be faithful to the original data's local structure while being simultaneously optimized for cluster separability. 

The first part of our study, which performs the initial, rough classification of Red Blood Cells (RBCs) and Platelets (PLTs), is directly inspired by the principles of the IDEC framework.
For use cases requiring more than just discriminative clustering, generative models offer a more sophisticated solution by learning the actual data distribution itself. The Variational Autoencoder (VAE), proposed by Kingma \& Welling, is a building block of the modern generative modeling \cite{Kingma2022}. The VAE is a learnable probabilistic mapping to a structured latent space from which new data points can be sampled. Building on this generative ability, Jiang et al. proposed Variational Deep Embedding (VaDE), a framework where a Gaussian Mixture Model (GMM) is employed as a prior in the VAE's latent space \cite{Jiang2017}. In VaDE, the data generation process is posed as first selecting a cluster from the GMM parameters end-to-end, VaDE learns a powerful generative model for each cluster. The second stage of our project, which aims to perform a more granular sub-classification of the RBC population, uses the VaDE approach for its ability to model complex, multi-modal distributions.

The practical relevance of this work is grounded in the technology of modern automated hematology analyzers, such as Aquarius series from Diatron~\cite{DiatronAquarius3}. 
These instruments generate a high-dimensional signal data that serves as the input for our algorithms. Our two-staged pipeline, which combines the discriminative power of IDEC with the generative ability of VaDE, is an approach to modeling this instrumental data in the low-resource scenario of sample-level reference counts. By recasting these high-level counts as soft probability targets, out algorithm has a semi-supervised nature, demonstrating an effective framework for applying state-of-the-art deep clustering techniques to real-world biomedical data challenges.

While this study focused on custom deep clustering architectures, future work will evaluate their performance relative to other hybrid or classical approaches using public datasets. Additionally, it is important to assess the pipeline's computational scalability and deployment feasibility for high-throughput lab settings. This includes evaluating the performance across various hardware configurations, as well as analyzing memory usage, and latency.

The current research examined a deep clustering algorithm for blood cell analysis via a two-stage pipeline with IDEC and VaDE models. The first stage utilized an IDEC-based model to detect red blood cells (RBCs) and platelets (PLTs) from a mixture of cell data. Although the model was trained in an semi-supervised manner (knowing only the number of cells per file), the integration of a regression head allowed for the prediction of sample-level RBC/PLT ratios, which helped guide the interpretation of clustering results. This method provided a way to assign per-cell labels that better matched known proportions from reference data.

A key observation during development was the importance of aligning model outputs with expected biological ratios. While semi-unsupervised clustering may often return arbitrary or non-symmetric groupings, post-processing with predicted ratios provided more interpretable and useful results. Using soft cluster probabilities and ranking them in order of likelihood was a handy workaround for having no ground truth cell-level annotations.

Stage two utilized a VaDE approach on the candidate set of cells verified as RBCs in an attempt to further separate into subtypes, such as pure RBCs, reticulocytes, and RBC clumps. The hierarchical architecture allowed the pipeline to be selective enough to process only relevant cells for subsequent clustering.

Two limitations of this study are: 
First, the algorithms were trained and tested on non-annotated cell-level data, which limits ultimate classification accuracy assessment. While the current approach uses sample-level counts and soft clustering, the addition of even partial cell-level labels, either manually annotated or synthetically generated from a developed simulator, could substantially improve model validation. Future work will concentrate on exploring ways of obtaining and using such labels.
Second, some design choices, such as the number of VaDE clusters or latent feature interpretation, remain somewhat heuristic and would be an interesting topic of investigation.

Despite these limitations, the work highlights the potential for deep clustering algorithms in biomedical applications. By blending representation learning with soft probabilistic assignments and post-processing, the pipeline demonstrated that valuable conclusions can actually be learned from unlabeled data. With improving annotation quality and greater accessibility, this approach could be a suitable scalable foundation for more advanced and accurate blood cell analysis systems.

A future direction would be the integration of explainable artificial intelligence (XAI) . 
In line with~\cite{Bhatia2023} and~\cite{Islam2024}, explainability could be achieved through methods such as feature attribution, latent space visualization or analysis of individual sample influence on cluster formation. Additionally, methods like SHAP (SHapley Additive exPlanations) \cite{SHAP2017} and LIME (Local Interpretable Model-Agnostic Explanations) \cite{LIME2016} could be explored to provide transparency in model decisions. 
Visualizing contributions of each feature to latent representations or clustering output would increase trust, while evaluation metrics for explainability~\cite{coroama2022evaluation} can contribute the more effective human-in-the-loop systems.  

As deep clustering methods continue to gain traction in biomedical data analysis, it becomes essential to enhance transparency, scalability, and clinical alignment. By addressing current limitations such as the lack of annotations, class imbalance, and manually selected parameters, and by integrating interpretability into the pipeline, future work aims to develop robust, efficient, and clinically trustworthy blood analysis systems that can generalize across different species and instruments.

\section{Conclusion}
This study presented a two-stage deep clustering pipeline aimed at detecting and classifying types of blood cells using unsupervised and semi-supervised learning methods. IDEC was used in the first stage to distinguish RBCs from platelets, while VaDE was used in the second stage to explore potential RBC subtypes.

The use of a latent space obtained from training an autoencoder with regression and clustering provided a generalizable framework for processing high-dimensional, unlabeled biological data. Although the algorithms themselves were based on relatively simple architectural components, they were able to produce cell-level classifications, generally in agreement with sample-level reference values.

This is especially valuable in real biomedical datasets, where class imbalance is common and often breaks standard models. The model’s strong RBC predictions, show its ability to generalize well. Just as important, processing over 100 large samples in about a minute highlights its practical efficiency. While platelet predictions remain more sensitive due to their low representation, the system is well-positioned to improve.
By capturing meaningful features in this latent space, the model can identify and quantify cell populations, even when it is faced with high class imbalance.

\begin{credits}
\subsubsection{\discintname}
The authors declare no competing interests.
\subsubsection{Acknowledgment} A. Groza is supported by the project "Romanian Hub for Artificial Intelligence-HRIA", Smart Growth, Digitization and Financial Instruments Program, MySMIS no. 334906.
\end{credits}

\end{document}